# Microstructure-Resolved Impedance Modeling of Solid-State Batteries

Kaniza Islam[a], Chengyin Wu[b], Noriko Katsube[a], Yanzhou Ji[b,*]

[a]*Department of Mechanical and Aerospace Engineering, The Ohio State University, Columbus, OH 43210, USA*

[b]*Department of Material Science Engineering, The Ohio State University, Columbus, OH 43210, USA*

## Abstract

Solid-state batteries (SSBs) with solid electrolytes (SEs) are attracting substantial investment from the automotive industry because of their potential to enable fast-charging and safer next-generation electric vehicles. The microstructure of the SE critically affects battery performance. Electrochemical impedance spectroscopy (EIS) is a powerful, non-destructive probe of charge-transfer and transport processes within SEs and across electrode/SE interfaces. However, quantitatively relating measured impedance spectra to underlying microstructural features remains an open modeling challenge. We present a two-component framework that combines microstructure generation with impedance prediction to predict microstructure-resolved impedance in a Li/$Li_6PS_5Cl$/Li symmetric cell. Spatially resolved $Li_6PS_5Cl$ microstructures are generated using phase-field sintering simulations, and impedance is computed for each fixed microstructure snapshot using an Ohmic conduction model incorporating Butler-Volmer kinetics and double-layer capacitance at the Li/SE interfaces. Across the sintering sequence, increasing porosity increases impedance, whereas improved inter-particle contact decreases it. When isolating the effects of the solid electrolyte interphase (SEI), we find that grain boundaries have only a minor effect on impedance, while SEI volume fraction and phase conductivity strongly influence the response. A multi-phase SEI significantly increases impedance. Finally, the same pipeline predicts impedance directly from experimental micrographs using AI-assisted pixel extraction, providing a pathway toward quantitative microstructure–impedance correlations based on real SE microstructures.

## 1. Introduction

Solid-state batteries (SSBs) are widely regarded as a leading candidate for next-generation electric vehicle (EV) energy storage, offering higher energy density and improved safety relative to conventional liquid-electrolyte lithium-ion cells, and the automotive industry is investing heavily in SSBs as a route to EVs that charge faster and operate more safely [1]. Realizing this promise, however, requires overcoming challenges at the solid-solid interface between the lithium-metal anode and the solid electrolyte (SE), where electro-chemo-mechanical instabilities can drive lithium dendrite growth, void formation, and interfacial contact loss, thereby increasing interfacial resistance and reducing round-trip energy efficiency [2,3]. Moreover, the microstructure of the SE evolves during SSB processing and operation, producing heterogeneous distributions of grain boundaries (GBs), residual pores, dead lithium, and solid-electrolyte-interphase (SEI) products that govern local $Li^+$ transport pathways and interfacial kinetics.

*Corresponding author. Email: ji.730@osu.edu

Electrochemical impedance spectroscopy (EIS) is widely used as an *in situ*, non-destructive diagnostic tool to probe charge-transfer kinetics in SSBs because it decomposes the cell response into frequency-dependent contributions that can, in principle, be mapped onto specific electrochemical processes, and distribution-of-relaxation-times (DRT) analysis has further been developed to deconvolve EIS spectra into contributions from distinct electrode processes [4]. Applying EIS/DRT analysis to Li/SE symmetric cells has shown that the impedance response is strongly influenced by the Li/SE interfacial microstructure, and that this sensitivity can be used to track interfacial degradation phenomena [5]. What EIS alone cannot do, however, is uniquely and quantitatively resolve the impedance contributions of individual microstructural feature such as GBs, pores, SEI layers, or isolated (dead) lithium domain from the overall measured spectrum, because Li/SE interfacial microstructures are dynamic and morphologically complex [6].

Physics-based modeling offers a way to address this limitation by explicitly resolving the electrochemical properties of individual phases and interfaces within the SE microstructure rather than treating impedance as an unresolved bulk property. However, existing modeling efforts have generally stopped short of coupling realistic, multi-feature two-dimensional (2D) or three-dimensional (3D) microstructures with full electro-chemo-mechanical physics. Prior impedance studies have attempted to incorporate microstructural information, such as correlating tomography or microscopy derived grain and pore features with electrochemical impedance spectra, but the impedance response itself is still interpreted using conventional equivalent-circuit curve fitting, with microstructure invoked only qualitatively to rationalize the fitted parameters [7,8].

This study aims to quantify how individual microstructural features of the SE such as grains, grain boundaries (GBs), pores, and solid electrolyte interphase (SEI) phases govern the impedance response of Li/SE symmetric cells. As a first, tractable step toward this goal, we adopt a simple Ohm's law based modeling framework rather than a fully coupled multiphysics description. The framework consists of two sequential components: microstructure generation and impedance prediction. Spatially resolved SE microstructures are obtained either from phase-field simulations of sintering and SEI formation or from segmentation of experimental micrographs, and each is represented as a phase map identifying the bulk SE, GBs, pores, and SEI/electrode interfaces (the SE domain in Fig. 1 is composed of these grain, GB, and pore regions). Each phase map is then imported into COMSOL Multiphysics, where every phase is assigned its corresponding ionic conductivity and other relevant transport parameters, and the impedance response is computed for that microstructure.

Although the underlying phase-field simulations are inherently dynamic, the impedance predictions reported here are computed for a series of fixed microstructure snapshots extracted from that evolution, rather than by simulating microstructure evolution and impedance concurrently. The impedance model describes electronic conduction in the lithium electrodes, ionic conduction in the SE and SEI phases, and interfacial charge transfer together with double-layer charging at the Li/SE or Li/SEI interface; mechanical deformation, stress, and concentration polarization are not included, making this a microstructure-resolved Ohmic conduction model. The novel contribution of this framework is the use of realistic, simulated or experimentally derived SE microstructures resolved down to individual grains, GBs, pores, and SEI phases, rather than idealized single-crystal geometries.

## 2. Modeling Approach

### *2.1 Model Geometry and Description*

The modeling approach couples simulated microstructures with physics-based impedance model. Different phase-field models are used to generate different classes of microstructural features relevant to SSBs which are described in Section 2.2. The specific cell configuration used in this study is a Li/$Li_6PS_5Cl$/Li symmetric cell (Figure 1). The SE layer is 700 μm thick and is sandwiched between two 50 μm-thick lithium metal electrodes. A small sinusoidal potential perturbation is applied at one electrode relative to an electrical ground at the other, consistent with a standard EIS measurement. The model is implemented in COMSOL Multiphysics and treats the SE as a single-ion conductor in which only the $Li^+$ carrier is mobile while the counter-ion sublattice remains fixed; Ohm's law describes charge transport in both the electrodes and the electrolyte, the interfacial current density is governed by the Butler-Volmer equation, and a double-layer capacitance is applied at both Li/SE interfaces.

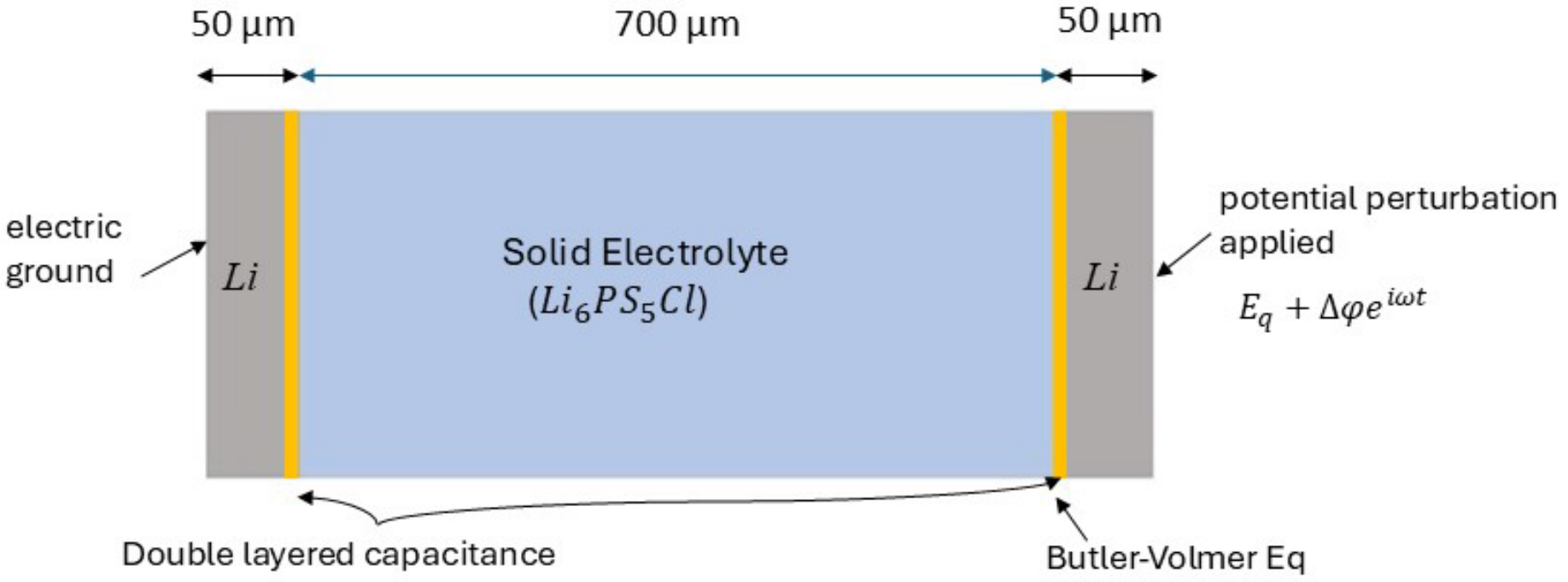

**Fig. 1.** Model geometry for the Li/$Li_6PS_5Cl$/Li symmetric-cell impedance simulation. A 700 μm SE layer is bounded by 50 μm lithium electrodes; a harmonic potential perturbation is applied at one electrode against an electrical ground at the other, with double-layer capacitance and Butler-Volmer kinetics imposed at both Li/SE interfaces.

In this study, each simulated or experimentally imaged microstructure is passed to the impedance-prediction model, where each phase and its associated features are assigned corresponding ionic transport properties. The model applies a harmonic voltage perturbation across the cell and solves the governing equations, including Butler-Volmer kinetics at the electrode/SE interfaces, ionic transport within the SE, and double-layer capacitance at the interfaces. This produces frequency-resolved Nyquist-type EIS predictions that reflect the specific microstructure supplied as input. Evaluating the model requires several microstructure-dependent material parameters, including ionic and electronic conductivity, and exchange current density, each of which can vary among the bulk grain, grain-boundary, pore, and SEI regions.

***Governing equations***

The charge transport in the SE domain follows Ohm's law.

$$\boldsymbol{i} = -\sigma \nabla \phi \tag{1}$$

where $\boldsymbol{i}$, $\sigma$, and $\phi$ represent the current density, ionic conductivity, and electric potential respectively. Conservation of electronic charge requires,

$$\nabla \cdot \boldsymbol{i} = 0 \tag{2}$$

The ionic conductivity $\sigma$ is phase-dependent and is assigned separately to the grain, grain-boundary, pore, and SEI phases.

***Li/SE interface***

The local current density $i_{local}$ at the Li/SE interface follows Butler-Volmer kinetics.

$$i_{local} = i_0 \left( e^{\frac{(1-\alpha)F\eta}{RT}} - e^{\frac{-\alpha F\eta}{RT}} \right) \tag{3}$$

where $i_0$ is the exchange current density, $\alpha = 0.5$ is the transfer coefficient, $F$ is the Faraday constant, $R$ is the gas constant, $T$ is the temperature and $\eta$ is the overpotential expressed as

$$\eta = \phi_{Li} - \phi_{SE} - E_{eq} \tag{4}$$

where $\phi_{Li}$ and $\phi_{SE}$ are the electric potential of Li metal and SE respectively and $E_{eq}$ is the equilibrium potential. The total current density is the summation of the interfacial Li-ion flux and the current density due to double-layer capacity,

$$\begin{aligned} i_{total} &= \sum i_{local} + i_{Cdl}, \\ i_{Cdl} &= C_{dl} \frac{\partial \eta}{\partial t} \end{aligned} \tag{5}$$

where $i_{Cdl}$ is the current density associated with the double-layer capacitive contribution.

***Impedance study***

To extract impedance, the left boundary is grounded, and a potential perturbation is applied at the right boundary, with the upper and lower boundaries electrically insulated. The area-specific impedance $Z$ is calculated as

$$Z(\omega) = \frac{\tilde{\phi}}{\tilde{i}_{total}} \tag{6}$$

where $\tilde{\phi}$=10 mV is the perturbation amplitude and $\omega = 2\pi f$ is the angular frequency. The impedance is simulated over a frequency range of 1 mHz to 1 MHz.

***2.2 Generation of Representative Microstructures using Phase-field Modeling***

Phase-field model is used to generate the porous polycrystalline microstructure of $Li_6PS_5Cl$ during sintering [9]. The microstructure is described by a conserved order parameter ρ distinguishing the pore $(\rho = 0)$ and the solid $(\rho = 0)$ phases, and a set of non-conserved order parameters $\{\eta_i\}$ distinguishing grains with different orientations. For example, in a system with two different grain orientations, $(\rho, \eta_1, \eta_2) = (0,0,0)$ represents the pore phase, $(\rho, \eta_1, \eta_2) = (1,1,0)$ represents the bulk of grain 1, and $(\rho, \eta_1, \eta_2) = (0,0,1)$) represents the bulk of grain 2; regions with $0 < \rho < 1$ represent the pore/solid interface while regions with $0 < \eta_1, \eta_2 < 1$ represents the grain boundary. The total free energy of the system is defined as

$$F = \int_V \left[ f(\rho, \{\eta_i\}) + \frac{1}{2}\kappa_\rho (\nabla \rho)^2 + \sum_i \frac{1}{2}\kappa_\eta (\nabla \eta_i)^2 \right] dv \tag{7}$$

Where $f(\rho, \{\eta_i\})$ is the nonequilibrium bulk chemical free energy density,

$$f(\rho, \{\eta_i\}) = A\rho^2(1-\rho)^2 + B\left[\rho^2 + 6(1-\rho)\sum_i \eta_i^2 - 4(2-\rho)\sum_i \eta_i^3 + 3\left(\sum_i \eta_i^2\right)^2\right] \tag{8}$$

With A and B being energy-related coefficients, the gradient terms $\frac{1}{2}\kappa_\rho(\nabla\rho)^2$ and $\sum_i \frac{1}{2}\kappa_\eta(\nabla\eta_i)^2$ accounts for interfacial energies due to inhomogeneous distributions of the order parameters at interfaces. The evolution equations of the order parameters are

$$\begin{gathered} \frac{\partial \rho}{\partial t} = \nabla \cdot M \nabla \frac{\delta F}{\delta \rho} = \nabla \cdot M \nabla \left( \frac{\partial f}{\partial \rho} - \kappa_\rho \nabla^2 \rho \right) \\ \frac{\partial \eta_i}{\partial t} = -L \frac{\delta F}{\delta \eta_i} = -L \left( \frac{\partial f}{\partial \eta_i} - \kappa_\eta \nabla^2 \eta_i \right) \end{gathered} \tag{9}$$

Where $L$ is a kinetic coefficient related to interface mobility, and $M$ is the effective diffusion mobility,

$$M = M_s h(\rho) + M_p[1 - h(\rho)] + M_{surf}\rho^2(1-\rho)^2 + M_{gb}\sum_{i,j\neq i} \eta_i^2 \eta_j^2 \tag{10}$$

Where $M_s$, $M_p$, $M_{surf}$ and $M_{gb}$ are diffusion mobilities in the bulk solid phase, bulk pore phase, solid/pore interface, and grain boundary, respectively. $h(\rho) = 6\rho^5 - 15\rho^4 + 10\rho^3$ is an interpolation function. For simplicity, we have neglected the effect of advection.

The simulations are performed in a 2-D 200*100 system with a grid spacing of 3.75 µm. 10 different grain orientations are considered. Initially, circular particles with an average radius of 60 µm are randomly introduced into the system; the number of particles is controlled to reach the desired porosity level. The governing equations are solved using Fourier spectral method with semi-implicit scheme [10]. The simulation parameters are listed in Table 1:

**Table 1.** Parameters for phase-field sintering simulations

| Parameters | Values |
|---|---|
| $A$ | 16 |
| $B$ | 1 |
| $\kappa_\rho$ | 4 |
| $\kappa_\eta$ | 4 |
| $M_s$ | 0.01 |
| $M_p$ | 0.001 |
| $M_{surf}$ | 1 |
| $M_{gb}$ | 0.1 |
| $L$ | 10 |

To further generate the SEI microstructure, we set $\rho$=1 and assign $\eta_1$, $\eta_2$, and $\eta_3$ to be the three SEI phases, while the rest non-conserved order parameters $(\eta_4 \sim \eta_{10})$ are assigned to be $Li_6PS_5Cl$ grains. This treatment effectively reduces the sintering model to a grain growth model. Initially, small circular particles of $\eta_1 \sim \eta_3$ are put randomly within the left 150 µm region, while large circular particles of $\eta_4 \sim \eta_{10}$ are put in the right region; the rest regions are hypothetical liquid phases with $(\eta_1, \eta_2, ..., \eta_{10}) = (0, 0, ..., 0)$. Simulations are performed until no hypothetical liquid regions left.

## 3. Results and Discussion

The EIS of the symmetric cell is simulated using the parameters tabulated in Table 2.

**Table 2.** Material properties used in the modelling.

| Parameters | Name | Symbol | Value | Unit | Ref |
|---|---|---|---|---|---|
| Electrochemical | Temperature | $T$ | 298 | K | |
| | Exchange current density | $i_0$ | 4.97 | $mA/cm^2$ | Exp |
| | Conductivity of $Li^+$ in SE ($Li_6PS_5Cl$) | $\sigma^+_{SE}$ | 0.319 | S/m | [11] |
| | Conductivity of $e^-$ of Li in Li-metal | $\sigma^+_{e^-}$ | $1.08x10^7$ | S/m | [12] |
| | Conductivity of $Li^+$ in $Li_2S$ | $\sigma^+_{Li_2S}$ | 0.001 | S/m | [13] |
| | Conductivity of $Li^+$ in $Li_3P$ | $\sigma^+_{Li_3P}$ | 0.01 | S/m | [13] |
| | Conductivity of $Li^+$ in LiCl | $\sigma^+_{LiCl}$ | $3.54x10^{-4}$ | S/m | [14] |
| | Conductivity of pores | $\sigma^+_{pore}$ | 0 | S/m | |
| | Double layer capacitance | $C_{dl}$ | 0.22 | $F/m^2$ | [7] |
| | Applied voltage | $\tilde{\phi}$ | 10 | mV | Exp |
| Geometry | Thickness of SE | $L_{SE}$ | 700 | μm | |
| | Thickness of Li-metal electrode | $L_a$ | 50 | μm | |

### *3.1. Impedance signature of sintering-induced microstructure evolution*

Sintering is the processing step that most directly determines the initial grain, grain-boundary, and pore structure of the SE, so its impedance signature is a natural first target for the microstructure-impedance framework. To isolate the effect of porosity, phase-field-generated microstructures were constructed over a 700×350 μm domain representing a cross-section of the SE at two nominal initial porosities, approximately ~12% and ~20%, alongside a fully dense single-crystal reference (Fig. 2a). The predicted impedance for these three cases shows that the semicircular arc shifts to progressively higher real-impedance values and grows in diameter as porosity increases from 0% (single crystal) to 12% to 20%, because pores lower the local $Li^+$ conductivity along their boundaries and through the residual void space, reducing the effective ionic cross-section available for transport.

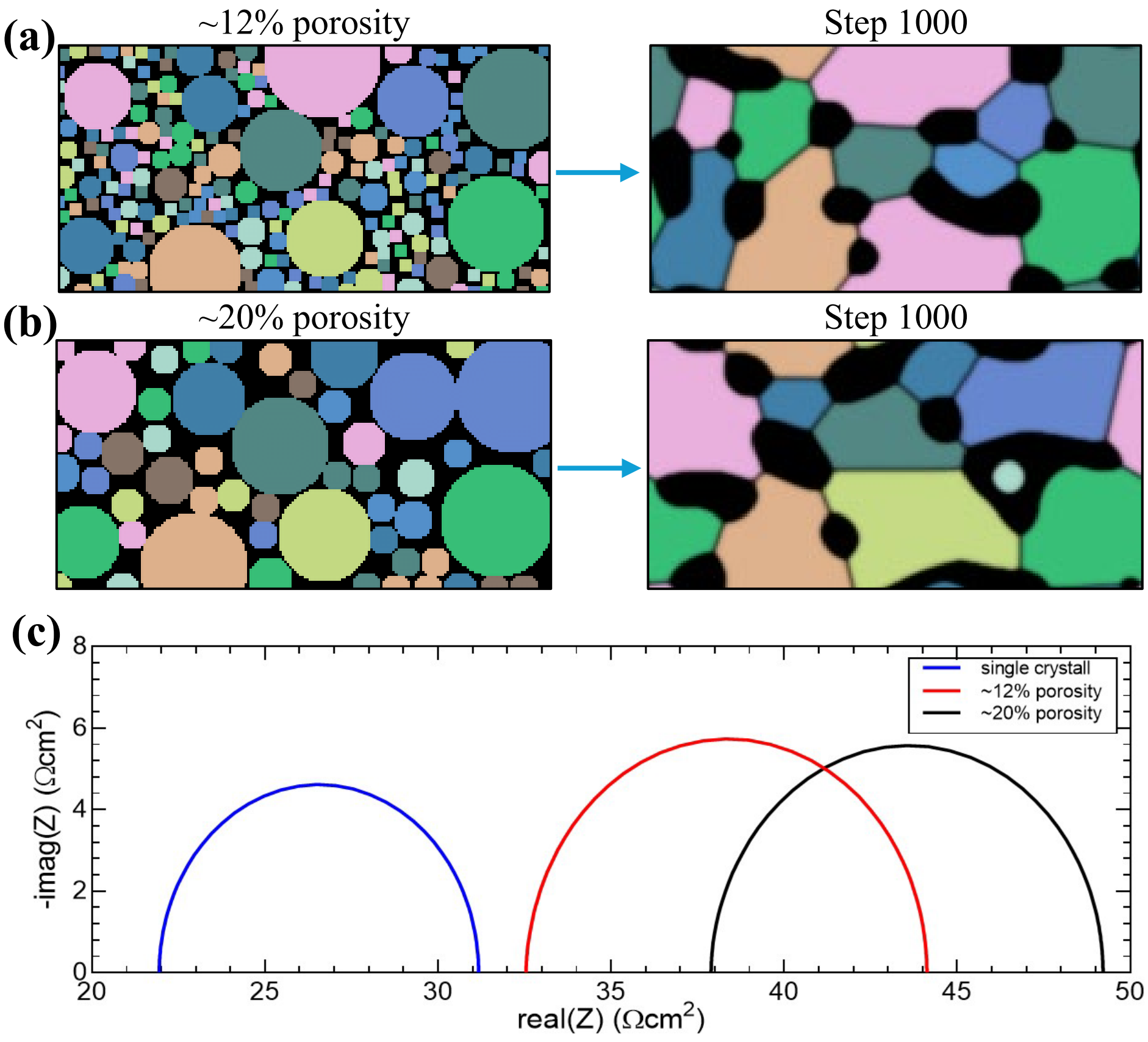

**Fig. 2.** Phase-field-generated SE microstructure evolution over a 700×350 μm domain and the corresponding predicted impedance. (a) Simulated microstructure with an initial porosity of ~12% at the initial and final stages of sintering. (b) Simulated microstructure with an initial porosity of ~20% at the initial and final stages of sintering. (c) Predicted impedance spectra for a single-crystal SE and the SE microstructures with porosities of ~12% and ~20% at step 1000 respectively.

To examine how impedance evolves as sintering proceeds, the ~20% porosity microstructure was advanced through five simulated sintering steps (step 1, 500, 1000, 10,000, and 20,000), tracking the progressive coarsening of grains and the coalescence of pores into fewer, larger voids (Fig. 3a). The porosity and GB maps in Fig. 3b and 3c use a 0/1 grayscale convention (0 = pore, 1 = dense SE in (b); 0 = grain interior, 1 = GB in (c)), and a 100 μm scale bar in Fig. 3a applies to all microstructure panels. Quantitative image analysis of these maps (Fig. 3e, f) shows that the porosity fraction does not decrease during this process: it rises modestly from about 19.5% at step 1 to approximately 22% by step ~5000 to 10,000 and remains approximately constant thereafter. The GB fraction, in contrast, rises sharply from near zero at step 1 reflecting the absence of a developed grain structure in the initial powder packing to a peak of about 4% by roughly step 2000 to 10,000, before decreasing to about 2.5–3% by step 20,000 as continued grain coarsening reduces the total grain-boundary. The corresponding grain-boundary network extracted at step 20,000 (Fig. 3c) is visibly sparser and more widely spaced than at earlier steps, consistent with this quantified decrease.

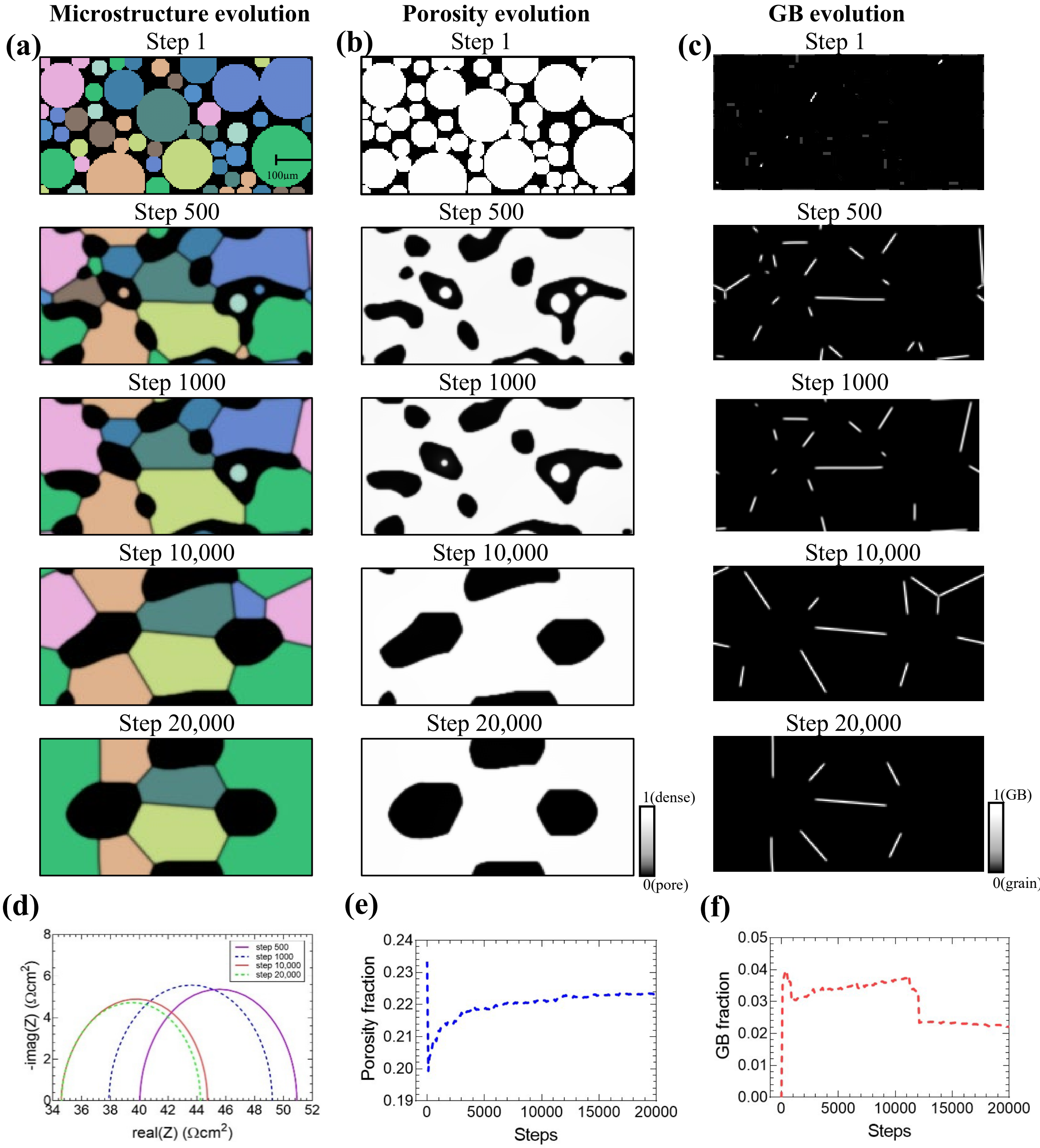

**Fig. 3.** Simulated microstructure evolution during sintering and the corresponding impedance response of the SE at ~20% porosity. (a) Simulated microstructure at steps 1, 500, 1000, 10,000 and 20,000 over a 700×350 μm domain. (b) Corresponding porosity evolution maps at each step. (c) Corresponding grain-boundary (GB) evolution maps at each step. (d) Predicted Nyquist impedance spectra at steps 1, 500, 1000, 10,000, and 20,000 showing the impedance change that accompanies the microstructural changes. (e) Porosity fraction as a function of simulation steps during sintering. (f) GB fraction as a function of simulation steps during sintering.

The predicted Nyquist impedance over this progression (Fig. 3d) shows the largest change over the early stage of sintering, with the semicircle contracting and shifting to lower real-impedance values as the GB network first forms and inter-particle contact improves; at later steps (10,000 and 20,000), the impedance response changes more gradually in Fig. 3e, f.

This progression in the Nyquist plot given the investigated microstructures can be described as follows. First, the high-frequency intercept corresponds to the bulk SE resistance; because grain size increases with continued sintering, the overall SE conductivity improves, and this intercept shifts to smaller real(Z) values. Second, the difference between the low- and high-frequency intercepts represents the combined interfacial polarization resistance of the two Li/SE interfaces. Due to the reorganization during the sintering, the combined interfacial polarization resistance of the Li/SE interfaces appears to become smaller. This will require further investigation on the Li/SE interfacial contact microstructure.

### *3.2. Effect of solid electrolyte interphase (SEI) formation on impedance*

In addition to porosity, the Li/SE interface is subject to chemical decomposition: argyrodite $Li_6PS_5Cl$ is known to react with metallic lithium to form a solid electrolyte interphase (SEI) composed of $Li_2S$, $Li_3P$, and LiCl, and this SEI layer is a poor $Li^+$ conductor that substantially increases interfacial impedance in experiments [15]. To quantify this effect within the present framework, a polycrystalline SE microstructure containing an explicit SEI layer at each Li/SE interface, together with grains and grain boundaries, was constructed (Fig. 4).

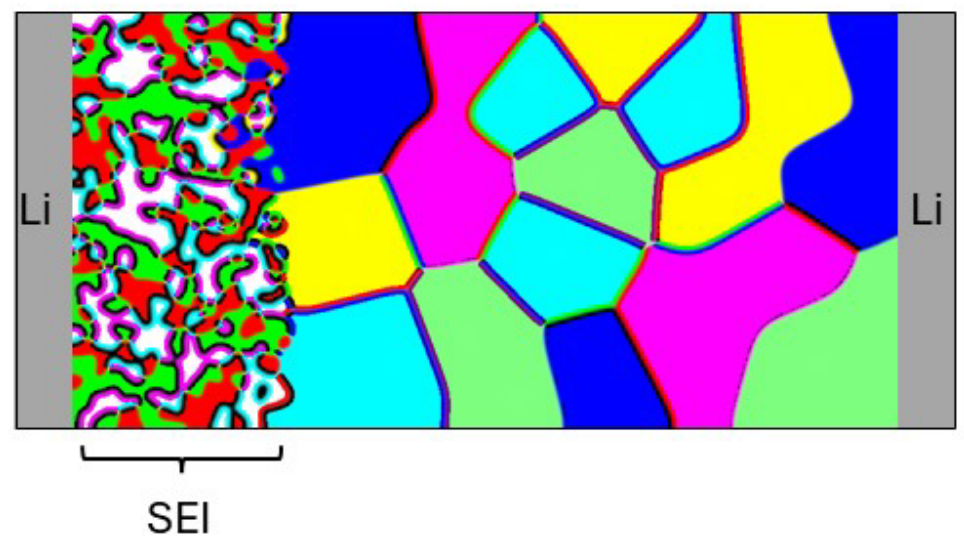

**Fig. 4.** Simulated polycrystalline SE microstructure containing grains, grain boundaries (GB), and an SEI layer at Li/SE interface.

To separate the contribution of grain boundaries from that of the SEI, a baseline comparison was first performed between a fully dense single-crystal SE and an SE containing only a grain-boundary network (no SEI), with GB $Li^+$ conductivity set to 0.01 times the bulk value (Fig. 5). The predicted Nyquist responses for these two cases are nearly indistinguishable, showing that a grain-boundary network alone in the absence of an SEI has only a marginal effect on the overall impedance, even when GB conductivity is two orders of magnitude lower than the bulk, given the investigated microstructures. This may establish grain boundaries as a comparatively minor contributor relative to the SEI effects examined next.

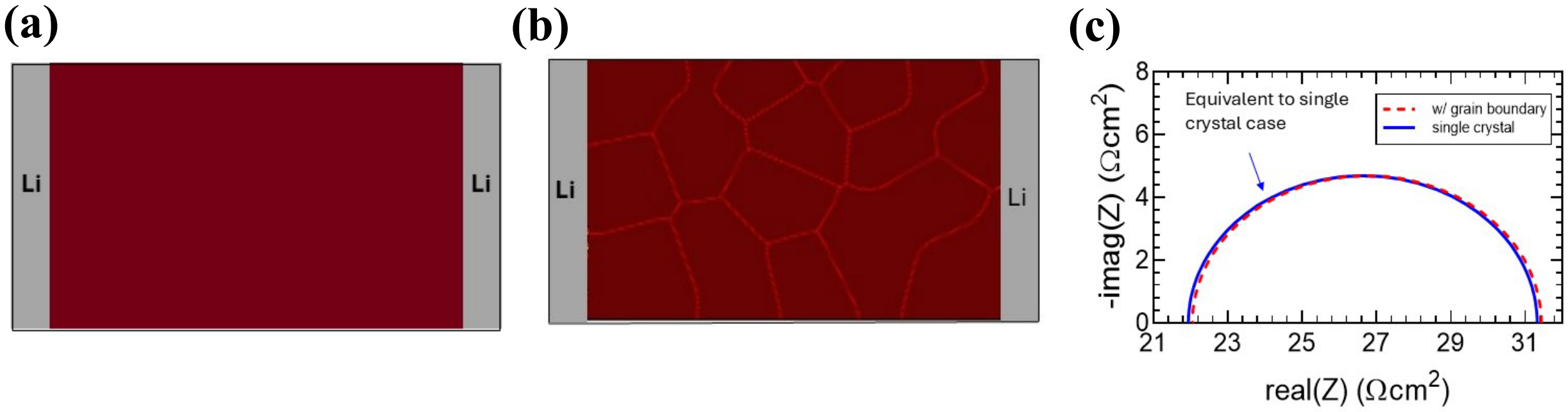

**Fig. 5.** Baseline comparison isolating the effect of grain boundaries in (a), SE containing a grain-boundary network with GB conductivity set to 0.01× the bulk value in (b), impedance showing the grain-boundary case is nearly equivalent to the single-crystal case in (c).

Building on this baseline, the volume fraction of the $Li_2S$ component of the SEI was then varied across three cases (Case 1: 8.8%, Case 2: 8%, Case 3: 7.6%) while keeping the grain/GB structure fixed (Fig. 6). The predicted Nyquist impedance shows a clear, monotonic sensitivity to SEI volume fraction: impedance increases with increasing Li2S content, confirming that even modest changes in the extent of interphase formation produce a measurable shift in the predicted EIS response relative to the grain-boundary-only baseline.

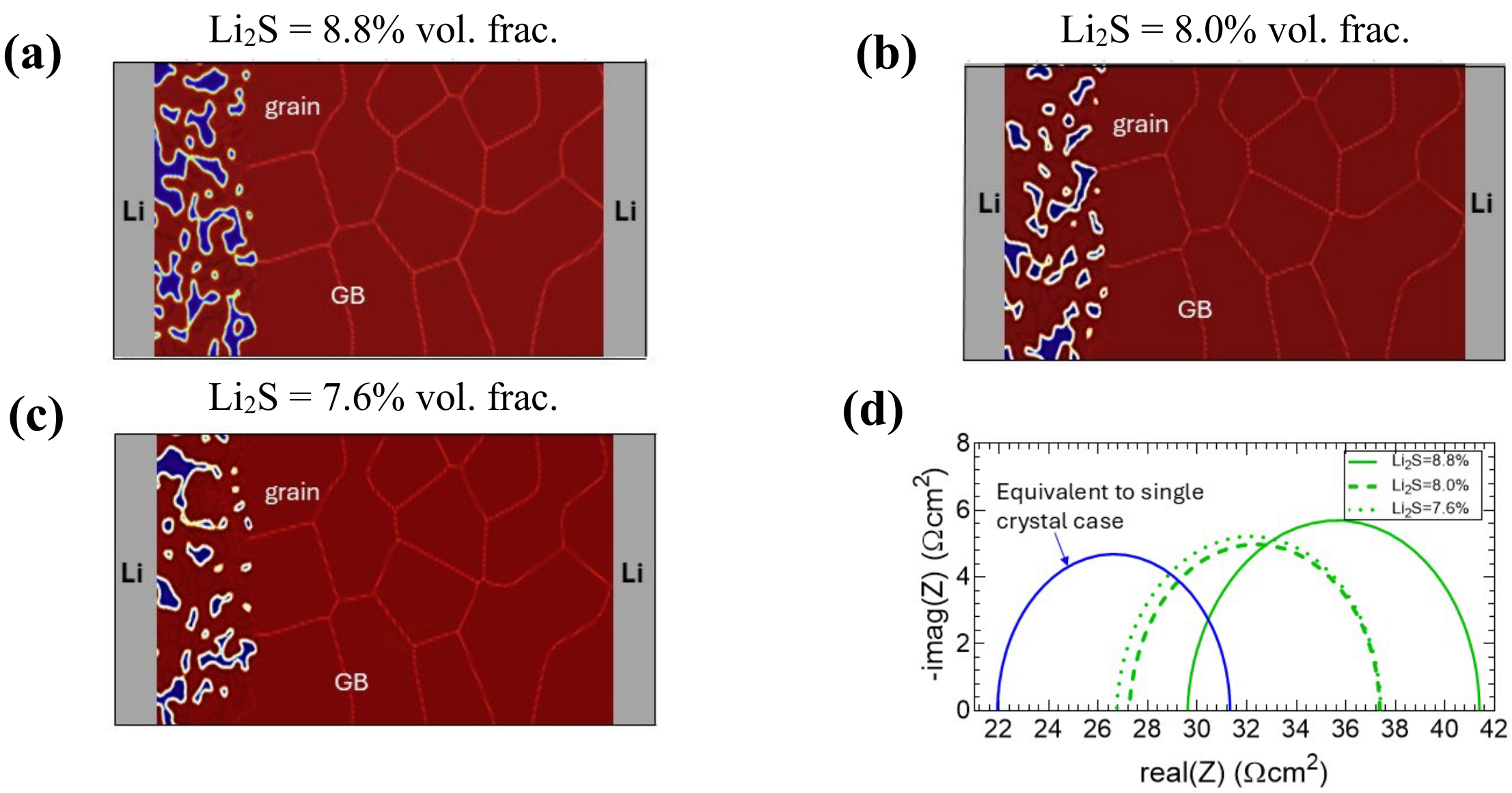

**Fig. 6.** Effect of SEI ($Li_2S$) volume fraction on predicted impedance. Simulated microstructures for (a) 8.8%, (b) 8%, and (c) 7.6% $Li_2S$ volume fraction. (d) Corresponding predicted impedance, showing the impedance increases with increasing $Li_2S$ content relative to the single-crystal baseline.

Because the SEI is a multi-phase layer, the individual conductivities of its three decomposition products $Li_2S$, $Li_3P$, and LiCl were then varied independently at a fixed SEI volume fraction of 8.8% (Fig. 7). Among the three assigned phase conductivities (one of which was set to a low value on the order of 0.01 S/m), the phase with the highest conductivity produces the smallest Nyquist arc, while the remaining, less conductive phases produce progressively larger arcs, consistent with the general trend that impedance increases as the conductivity of the SEI phase decreases. This is qualitatively consistent with reports that $Li_3P$ is comparatively more favorable for $Li^+$ transport than $Li_2S$ and LiCl within a mixed SEI, and it demonstrates that the framework can resolve impedance sensitivity to the transport properties of individual SEI phases, not just their combined presence.

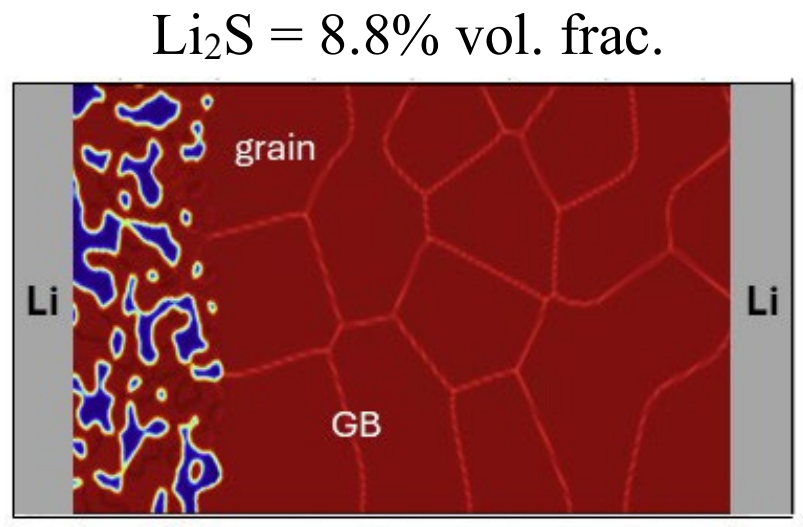

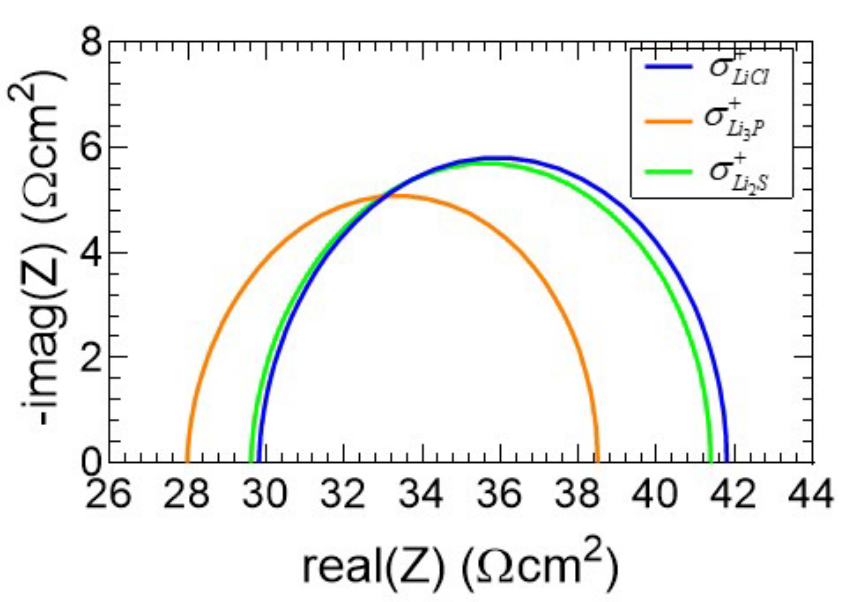

**Fig. 7.** Effect of individual SEI-phase conductivity on predicted impedance where SEI volume fraction is 8.8%. (a) simulated microstructure. (b) Predicted impedance obtained by independently assigning the conductivity of LiCl, $Li_3P$, and $Li_2S$, showing that impedance increases as the assigned phase conductivity decreases.

Finally, all three SEI decomposition products ($Li_2S$, $Li_3P$, and LiCl) were considered together, alongside the grain and grain-boundary structure, in a single simulation (Fig. 8). Relative to the grain-boundary-only baseline (real(Z) ≈ 22–32 Ωcm², Fig. 5), the combined multi-phase SEI shifts the predicted impedance to real(Z) ≈ 586–599 Ωcm². Much of this change appears as an increased ohmic offset, and it can become the primary microstructural feature responsible for degrading Li/SE interfacial performance during interphase formation. Note that the SEI layer thickness used here is exaggerated relative to physically realistic values, so as to isolate and clearly illustrate the effect of SEI composition on impedance.

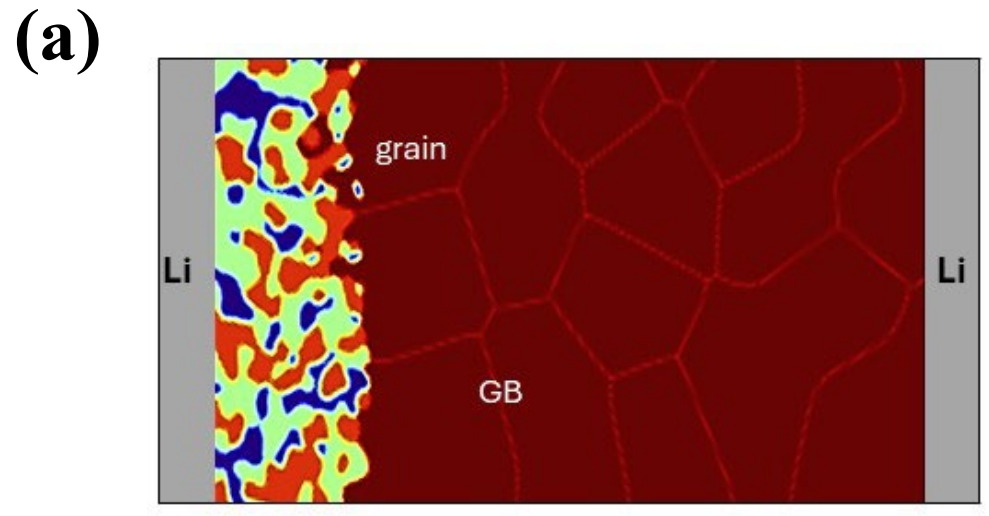

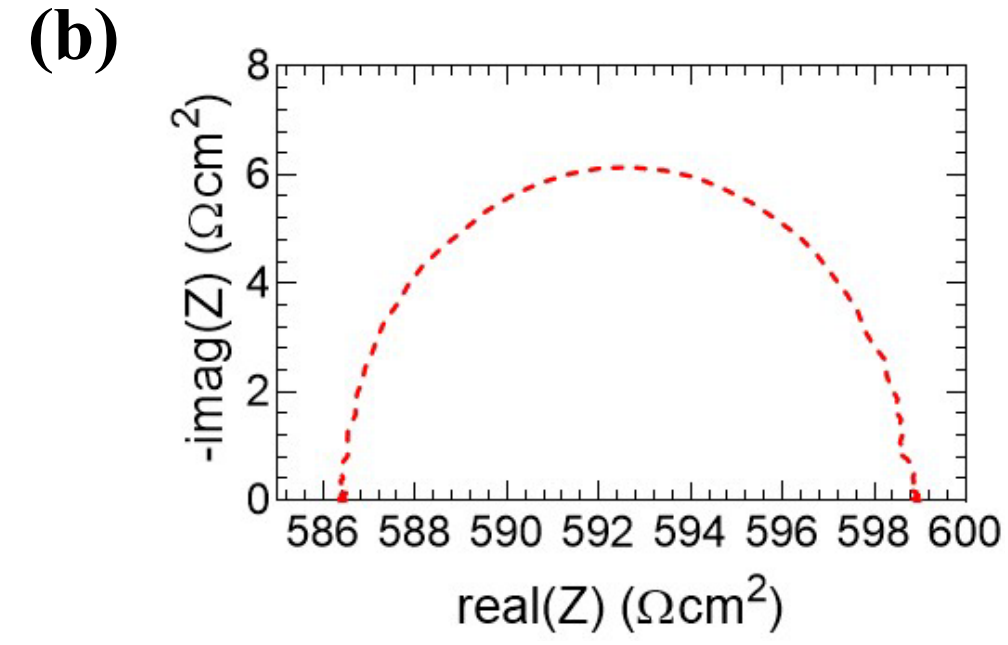

**Fig. 8.** Combined effect of a multi-phase SEI ($Li_2S$, $Li_3P$, and LiCl) together with grains and grain boundaries. (a) Simulated microstructure, with SEI phases indicated (blue: LiCl; green: $Li_2S$; orange: $Li_3P$). (b) Predicted impedance.

### 3.3. Extension to experimentally imaged microstructures

A key requirement for the framework is that it generalizes beyond simulated microstructures to microstructures obtained directly from experimental micrographs, since these are ultimately what characterize as-fabricated SE layers. To test this, a cross-sectional scanning electron microscopy (SEM) image of a $Li/Li_6PS_5Cl$ (LG-LPSCl) interface from the literature was used as the microstructure input (Fig. 9) [16]. Rather than manually segmenting the image, an AI-assisted image-processing tool (ChatGPT) was used to convert the raw pixel intensities of the highlighted region of the micrograph into discretized phase (grain/grain-boundary/pore) information suitable for the impedance solver.

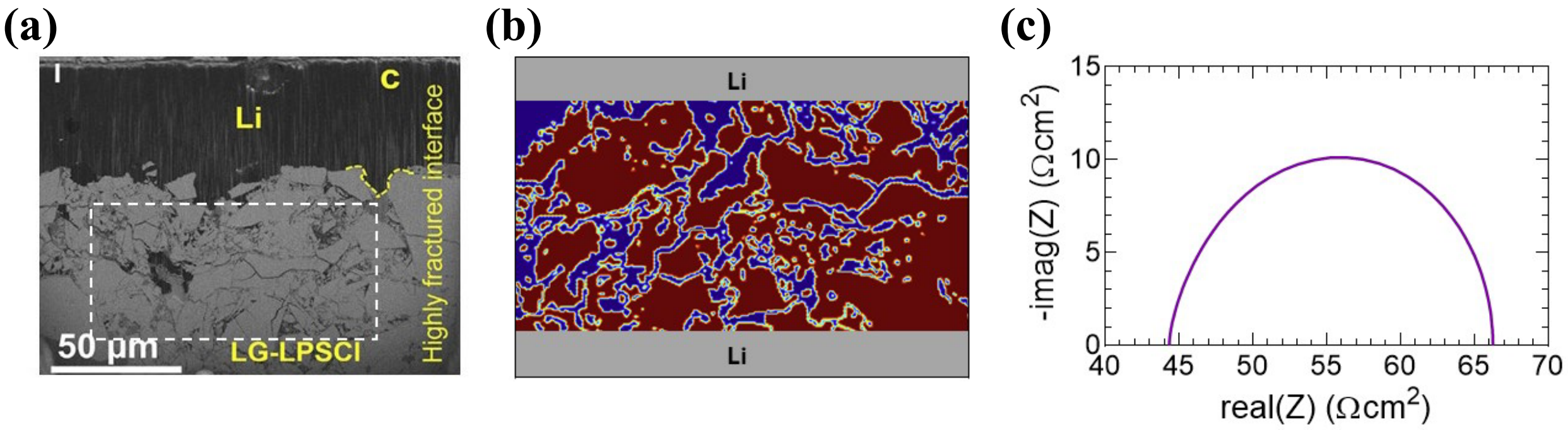

**Fig. 9.** Extraction of pixel-level microstructure information to predict impedance directly from an experimental micrograph. (a) Cross-sectional SEM image of a $Li_6PS_5Cl$ SE adjacent to a Li electrode, from the literature [10], with the region used for microstructure extraction outlined. (b) The extracted phase map obtained via AI-assisted pixel classification, and (c) the resulting predicted Nyquist impedance computed from that extracted microstructure.

The resulting predicted impedance spectrum (Fig. 9c) has the same qualitative semicircular Nyquist shape obtained from the phase-field-generated sintering microstructures in Section 3.1, but with a magnitude and offset set by the specific grain, grain-boundary, and pore arrangement present in the literature micrograph. This demonstrates that the pixel-extraction step can serve as a bridge between purely computational microstructure generation and real, imaged SE microstructures, allowing the impedance-prediction module to be applied directly to experimental characterization data without requiring a separate phase-field simulation of that specific sample.

## 4. Conclusion

This work builds a modeling framework that links microstructures of $Li_6PS_5Cl$ (taken from phase-field simulations) to an electrochemical impedance model based on Ohm's law, to predict the EIS response of $Li/Li_6PS_5Cl/Li$ symmetric cells.

For the phase-field-generated porous microstructures, increasing porosity increases the calculated effective ohmic resistance by reducing the connected cross-sectional area available for ionic transport. During the simulated microstructural evolutions, changes in pore morphology, grain structure, grain-boundary fraction and interparticle connectivity collectively alter the impedance response. Within the microstructures examined here, increase in grain size appears to improve the overall SE conductivity.

For SEI formation, the framework shows that grain boundaries alone raise impedance only slightly given the investigated microstructure. Impedance instead scales with the volume fraction of SEI and the conductivity of each SEI phase. A multi-phase SEI ($Li_2S$, $Li_3P$, LiCl) increases impedance compared to a pristine grain-boundary network. This may identify SEI formation, not grain boundaries or moderate porosity, as the main driver of degradation.

The same approach was also applied to a real SEM image using AI-assisted pixel extraction. This shows the framework is not limited to simulated microstructures and can, in principle, predict impedance directly from experimental micrographs as well.

Together, these results show that microstructure-resolved impedance modeling can quantitatively connect the processing and degradation state of a solid electrolyte (sintering degree, porosity, grain-boundary structure, and SEI composition) to its impedance signature. This is an important step toward using EIS not just to detect degradation, but to identify its specific microstructural cause, and toward guiding microstructure and interphase design (for example, targeted sintering protocols or SEI-suppressing coatings) to engineer solid electrolytes with lower impedance.

## Acknowledgments

K.I. acknowledges the financial support of NSF National Research Traineeship Program, The Ohio State EmPOWERment Program (Grant # 1922666). Y.J. acknowledges the Ohio State University Materials Research Seed Grant Program, funded by the Center for Emergent Materials, an NSF-MRSEC, grant DMR-2011876, the Center for Exploration of Novel Complex Materials, and the Institute for Materials and Materials Research. The authors also acknowledge Smart Vehicle Concepts Center (www.SmartVehicleCenter.org), a graduated National Science Foundation Industry–University Cooperative Research Center initially established under Grant NSF IIP 1738723.